\documentclass[prd,letterpaper,twocolumn,showpacs,showkeys,lengthcheck,
               floatfix,nofootinbib,preprintnumbers,superscriptaddress]{revtex4-1}

\usepackage{amsmath,amssymb,amsfonts,graphicx}
\usepackage{bm}
\usepackage{slashed}
\usepackage[caption=false]{subfig}
\usepackage[svgnames]{xcolor}
\usepackage[english]{babel}
\usepackage{blindtext}
\usepackage{microtype}
\usepackage{tikz}
\usepackage{dcolumn}
\usepackage{multirow}
\usepackage{ifpdf}
\usepackage[colorlinks=true]{hyperref}

\ifpdf
  \makeatletter
  \g@addto@macro\Gin@extensions{,.eps}
  \makeatother
\fi

\newcommand{\sh}[1]{\slashed{#1}}
\newcommand{\vect}[1]{\boldsymbol{#1}}

\def\g{\gamma}

\def\l{\lambda}
\def\ve{\varepsilon}
\def\vp{\varphi}
\def\s{\sigma}
\def\o{\omega}

\def\L{\Lambda}

\def\hs{\hspace}

\def\no{\nonumber}

\def\lf{\left}
\def\rg{\right}
\def\la{\langle}
\def\ra{\rangle}

\def\ua{\uparrow}
\def\da{\downarrow}

\graphicspath{{.}{figures/}} 
\newcommand{\eq}[1]{Eq.~(\ref{#1})}

\begin{document}

\preprint{NT@UW-12-07} 

\title{Nucleon form factors and spin content in a quark-diquark model with a pion cloud}

\author{Ian C. Clo\"{e}t}
\affiliation{Department of Physics, University of Washington, Seattle, WA 98195-1560, USA}
\affiliation{The Special Research Centre for the Subatomic Structure of Matter, \\ 
School of Chemistry and Physics
University of Adelaide, Adelaide SA 5005, Australia}

\author{Gerald A. Miller}
\affiliation{Department of Physics, University of Washington, Seattle, WA 98195-1560, USA}

\begin{abstract}
We propose a new  model of the nucleon  in which  quark-diquark configurations 
immersed in a pion cloud are treated in a manner consistent with Poincar\'{e} invariance.  
With suitably chosen parameters, the computations employing this model reproduce the
measured electromagnetic form factors and the  quark-spin contribution to the total 
nucleon angular momentum.
\end{abstract}


\maketitle

\section{Introduction}

The nucleon is the lightest baryon and its mass dominates the nucleus, which is the heart of the atom.
Quantum chromodynamics tells us that the nucleon is a complex system composed of three valence quarks
and an undefined number of quark-antiquark pairs and gluons.  Deep inelastic scattering measurements 
have demonstrated that the sum of the spins of the quarks do not add up to the total angular momentum of 
the nucleon~\cite{Ashman:1987hv}. This puzzle has been the subject of a tremendous amount of experimental and theoretical 
investigation. Another probe of the structure of the nucleon is  elastic electron-nucleon scattering. 
Measurements made at Jefferson Lab have shown that the proton form factor ratio, $G_{Ep}(Q^2)/G_{Mp}(Q^2)$, 
decreases as the value of 
$Q^2$ is increased above about 1 GeV$^2$. This important discovery renewed interest in the structure of the nucleon.

The present paper is devoted to answering a  simple question: can a model of the nucleon which consists of 
three valence quarks and a pion cloud, constrained by Poincar\'{e} invariance, describe the existing data for elastic
electromagnetic form factors, while properly accounting for the  small fraction of the proton total angular 
momentum carried by
the quarks.  Recent  work indicates that the successful construction of such a model should be 
possible~\cite{Thomas:2008ga},
provided the model quark wave functions have suitable properties.

The challenge  of understanding nucleon elastic form factors has been taken up by many, for example, see the 
review~\cite{Arrington:2011kb} and Ref.~\cite{Cloet:2008re}. Here we follow only one particular line of reasoning. 
The light-front model of Ref.~\cite{Frank:1995pv}, with three constituent quarks, was used to predict the fall of the 
ratio $G_{Ep}(Q^2)/G_{Mp}(Q^2)$.  The effects of the pion cloud were later 
included~\cite{Miller:2002ig,Matevosyan:2005bp}, and this led to a 
reasonably accurate  description of all four electromagnetic form factors. However, the quarks
in the bare nucleon carry about 75\% of the total angular momentum of the nucleon,  which is too 
large to reproduce the measured value of approximately 30\%. 
Furthermore, the computed ratio $G_{Ep}(Q^2)/G_{Mp}(Q^2)$ falls a little too rapidly with increasing values of $Q^2$, and 
the results were not completely consistent with the  detailed flavor decomposition of the empirical form 
factors~\cite{Cates:2011pz}.
This earlier work on the proton form factors was carried out with a very simple three-quark wave 
function.
In the present work we use a more sophisticated wave function, consisting of a quark-scalar-diquark term and a 
quark-axial-vector-diquark term, with two invariant forms for each term. 

The plan of the paper is as follows: Sect.~\ref{sec:formal} is devoted to a complete description of the model, 
including the light-front wave function (LFWF) and the addition of the 
pion cloud, along with the formalism necessary to compute observable quantities. 
The parameters of the model are discussed in Sect.~\ref{sec:res}, where they are varied to describe the existing 
data for nucleon electromagnetic form factors.
The choice of parameters completes the definition of the model.  
The model is tested in Sect.~\ref{sec:spin} by computing the quark contribution to the nucleon spin and
Sect.~\ref{sec:sum} is reserved for a summary and discussion.

\section{A Covariant Light-Front Model for the Nucleon}
\label{sec:formal}
The basic model is that the valence quarks, represented by quark-diquark combinations with the 
quantum numbers of the nucleon, are immersed in a cloud of pions. 
The motivation for this idea is obvious. We know that the nucleon is made of quarks and that
there is a long range interaction between nucleons mediated by the exchange of a single pion. 
However, a pion emitted by a nucleon can be absorbed by the same nucleon, so each nucleon has 
a pion cloud. The low mass of the pion is the reason for singling it out as the only meson to be 
treated separately as a cloud~\cite{Theberge:1980ye}.  As we shall see, including the pion cloud 
leads to a significant reduction in the fraction of the nucleons total angular momentum carried
by the quark spin, and this is consistent with previous findings~\cite{Thomas:2008ga,Myhrer:2007cf}.
We use the light-front representation of the nucleon wave function~\cite{Brodsky:1997de} to 
guarantee that the observable quantities have the appropriate properties under Lorentz transformations. 
The remainder of this section details how this is done.

In general, the light-front wave function (LFWF) of a hadron with spin projection 
$J_z=\pm\tfrac{1}{2}$ is represented by the 
function $\Psi^{J_z}_{\l_1,\ldots,\l_n}(x_i,\vect{k}_{\perp i})$~\cite{Brodsky:1997de}, where
\begin{align}
k_i = \lf(k_i^+,k_i^-,\vect{k}_{\perp i}\rg) 
= \lf(x_i\,p^+,\tfrac{\vect{k}_{\perp i}^2 + m_i^2}{x_i\,p^+},\vect{k}_{\perp i}\rg),
\end{align}
specifies the 4-momentum of each constituent and $\l_i$ specifies its light-front
helicity in the $z$-direction. The light-front momentum fractions, 
$x_i = \frac{k^+_i}{p^+}$, are all positive and satisfy $\sum_i x_i = 1$. The scalar parts 
of the LFWF are functions of the Lorentz invariant quantities $x_i$ and the invariant mass squared, $M_0^2$,
given by
\begin{align}
M_0^2 = \sum_i^n\,\frac{\vect{k}_{\perp i}^2 + m_i^2}{x_i} = \lf(\sideset{}{_i}\sum k_i\rg)^2,
\end{align}
where $m_i$ is the mass of each nucleon constituent.

For a nucleon that consists of two constituents, in our case a quark and a diquark, 
the nucleon Fock state can be expressed as
\begin{multline}
\lf\lvert p^+,\vect{p}_{\perp}=0,\,\l \rg\ra =
\int \frac{dx\,d^2k_\perp}{16\pi^3\sqrt{x(1-x)}} \\ 
\sum_{\l_q,\l_a}\, 
\Psi^{\l}_{\l_q \l_a}(x,\vect{k}_{\perp}) \lf\lvert xp^+,\vect{k}_{\perp},\l_q,\l_a \rg\ra,
\label{eq:lfwf}
\end{multline}
where $\Psi^{\l}_{\l_q \l_a}(x,\vect{k}_{\perp})$ is the LFWF that describes the interaction of
a quark and a diquark to form a nucleon. We have chosen a frame where the transverse
momentum of the nucleon is zero, and the helicities of the quark and diquark states are
labeled by $\l_q$ and $\l_a$, respectively. The two particle Fock-state ket in Eq.~\eqref{eq:lfwf} is 
defined by
\begin{multline}
\lf\lvert xp^+,\vect{k}_{\perp},\l_q,\l_a \rg\ra \equiv 
\big\lvert k_1^+ = x\,p^+, k_2^+ = (1-x)p^+, \\
k_{1\perp} = k_{\perp}, k_{2\perp} = -k_{\perp}; \l_q,\l_a \bigr\ra.
\end{multline}

In this work we make the quark-diquark approximation for the LFWF of the nucleon, where we include
both scalar and axial-vector diquark correlations. The LFWF then takes the form
\begin{multline}
\Phi^{\l}_{\l_q\,\l_a}(k,p) = \bar{u}_q(k,\l_q)\lf[\varphi^s_1 + \frac{M\,\sh{\o}}{\o\cdot p}\varphi^s_2\rg]u(p,\l) \\
  + \bar{u}_q(k,\l_q)\,\ve^*_\nu(q,\l_a)\g^\nu\g_5\lf[\varphi^a_1 + \frac{M\,\sh{\o}}{\o\cdot p}\varphi^a_2\rg]u(p,\l),
\label{eq:nucleon_LFWF1a}
\end{multline}
where the first term represents correlations in the quark--scalar-diquark channel 
and the second quark--axial-vector-diquark correlations. The variables
$k,~q,~p$ are respectively the quark, diquark and nucleon momentum, where $p=k+q$
and $M$ is the nucleon mass.
The quark and nucleon
spinors are represented by $u_q(k,\l_q)$ and $u(p,\l)$, respectively, and $\ve^\mu(q,\l_a)$ is the usual
spin-one polarization vector, representing the spin-one axial-vector diquark.
The interaction of the quark with the diquark, in each diquark channel, is encapsulated by two scalar
functions, namely $\varphi_1$ and $\varphi_2$. We choose the $\vp_1$ and $\vp_2$ scalar functions to 
have the form
\begin{align}
\varphi_1  = \frac{1}{\lf(M_0^2 + \beta^2\rg)^\g}, \qquad
\varphi_2  = c\,\frac{\lf(M_0  - M\rg)}{2\,M}\varphi_1.
\label{phi}
\end{align}
This choice is motivated by the success of earlier work described in Ref.~\cite{Brodsky:2003pw}.

The wave function given in Eq.~\eqref{eq:nucleon_LFWF1a} is defined at the light-front plane $\o\cdot x = \s$, where $\o$ is
a light-like vector. For a stationary state we can consider a fixed light-cone time 
and set $\s=0$. The usual choice for the quantization direction is $\o=\lf(1,0,0,-1\rg)$,
so the nucleon wave function becomes
\begin{align}
\Phi^{\l}_{\l_q\,\l_a}(k,p) &= \phi^{\l}_{\l_q}(k,p) + \phi^{\l}_{\l_q\,\l_a}(k,p), \no \\
&= \bar{u}_q(k,\l_q)\lf[\varphi_1^s + \frac{M}{p^+}\,\g^+\,\varphi_2^s\rg]u(p,\l) \no \\
&\hs{-20mm}
+ \bar{u}_q(k,\l_q)\,\ve^*_\nu(q,\l_a)\g^\nu\g_5\lf[\varphi_1^a + \frac{M}{p^+}\,\g^+\,\varphi_2^a\rg]u(p,\l),
\label{eq:nucleon_LFWF2a}
\end{align}
where $\phi^{\l}_{\l_q}(k,p)$ represents the quark--scalar-diquark component and $\phi^{\l}_{\l_q\,\l_a}(k,p)$
the quark--axial-vector-diquark component of the nucleon LFWF.
The above wave function contains only the spin couplings, therefore to fully define the
model we also need the flavor couplings. The flavor wave function of the proton is given by
\begin{align}
\lvert p \ra = \frac{1}{\sqrt{2}} \lvert u\,S \ra + \frac{1}{\sqrt{6}} \lvert u\,T_0 \ra
                                                  - \frac{1}{\sqrt{3}} \lvert d\,T_1 \ra, 
\label{flave}
\end{align}
where $S$ is the flavor singlet state and $T$ the flavor triplet and therefore we obtain a
symmetric spin-flavor wavefunction.

\subsection{Bare nucleon form factors}
\label{sec:bare}
For on-shell initial and final nucleon states, the Dirac and Pauli electromagnetic form factors
of the nucleon are defined via the matrix element decomposition
\begin{multline}
\la p',\,\l' \lf\lvert J_{\text{em}}^\mu\rg\rvert p,\,\l \ra = \\
\bar{u}(p',\l')\lf[\g^\mu\,F_1(Q^2) + \frac{i\s^{\mu\nu}q_\nu}{2\,M}\,F_2(Q^2)\rg]u(p,\l),
\end{multline}
where $M$ is the nucleon mass and $Q^2 = -q^2$, where $q$ is the 4-momentum transfer.
We choose to work in the Drell-Yan-West frame, where the light-front momentum decompositions
of the relevant 4-vectors are
\begin{align}
q &= \lf(q^+,q^-,\vect{q}_\perp\rg) = \lf(0,\tfrac{Q^2}{p^+},\vect{q}_\perp\rg), \\
p &= \lf(p^+,p^-,\vect{p}_\perp\rg) = \lf(p^+,\tfrac{M^2}{p^+},\vect{0}_\perp\rg), 
\end{align}
so that  $q^2 = -2\,p\cdot q = -\vect{q}_\perp^2 = -Q^2$.
With this choice the Dirac and Pauli from factors are identified with the helicity-conserving and
helicity-flip matrix elements of the plus-component of the electromagnetic current, that is
\begin{align}
\label{eq:f1}
F_1(Q^2) &= \frac{1}{2\,p^+}\,\lf\la p',\ua \lf\vert J_{\text{em}}^+ \rg\lvert p,\ua \rg\ra, \no \\
         &= \frac{1}{2\,p^+}\,\lf\la p',\da \lf\vert J_{\text{em}}^+ \rg\lvert p,\da \rg\ra, \displaybreak[2] \\[1.0ex]
\label{eq:f2}
F_2(Q^2) &= \frac{-2\,M}{(q^1-iq^2)}\,\frac{1}{2\,p^+}\lf\la p',\ua \lf\vert J_{\text{em}}^+ \rg\lvert p,\da \rg\ra,\no  \\
        &= \frac{2\,M}{(q^1+iq^2)}\,\frac{1}{2\,p^+}\lf\la p',\da \lf\vert J_{\text{em}}^+ \rg\lvert p,\ua \rg\ra.
\end{align}
To determined the nucleon form factors we must therefore compute the above matrix elements of the 
$J^+$ component of the electromagnetic current.

Using Eq.~\eqref{eq:lfwf} and the matrix element definitions of the nucleon form factors given 
in Eqs.~\eqref{eq:f1} and \eqref{eq:f2}, it is clear that the nucleon form factors are given by
\begin{align}
F_1(Q^2) &= \int \frac{dx\,d^2 k_\perp}{\sqrt{x(1-x)}} \no \\
&\hs{-3mm}  \times           
\sum_{\l_q,\,\l_a=+,-} \ \Psi^{\ua *}_{\l_q,\,\l_a}(x,\vect{k}'_{\perp}) \Psi^{\ua}_{\l_q,\,\l_a}(x,\vect{k}_{\perp}),  \\
F_2(Q^2) &= \frac{2\,M}{q^1 + iq^2}\, \int \frac{dx\,d^2 k_\perp}{\sqrt{x(1-x)}}\no \\
&\hs{-3mm} \times           
\sum_{\l_q,\,\l_a=+,-} \ \Psi^{\da *}_{\l_q,\,\l_a}(x,\vect{k}'_{\perp}) \Psi^{\ua}_{\l_q,\,\l_a}(x,\vect{k}_{\perp}).
\end{align}
The helicity components of the LFWFs for scalar and axial-vector diquarks are defined via
\begin{align}
\label{psis}
\psi^{\l}_{\l_q}(x,\vect{k}_\perp) &= \frac{1}{\sqrt{x(1-x)}}\,\phi^{\l}_{\l_q}(k,p), \\
\label{psiav}
\psi^{\l}_{\l_q\,\l_a}(x,\vect{k}_\perp) &= \frac{1}{\sqrt{x(1-x)}}\,\phi^{\l}_{\l_q\,\l_a}(k,p),
\end{align}
and for convenience we define the scalar functions 
\begin{align}
\label{eq:fs1}
f^s_{1}(Q^2) &= \int \frac{dx\,d^2k_\perp}{16\pi^3}         
\sum_{\l_q=+,-}  \psi^{\ua *}_{\l_q}(x,\vect{k}'_{\perp}) \psi^{\ua}_{\l_q}(x,\vect{k}_{\perp}), \\
f^a_{1}(Q^2) &= \int \frac{dx\,d^2k_\perp}{16\pi^3}  \no \\
&\hs{0mm}  \times 
\sum_{\l_q,\,\l_a=+,-}         
\psi^{\ua *}_{\l_q,\,\l_a}(x,\vect{k}'_{\perp}) \psi^{\ua}_{\l_q,\,\l_a}(x,\vect{k}_{\perp}),\\[1.0ex]
f^s_{2}(Q^2) &= \frac{2M}{q^1+i\,q^2}\int \frac{dx\,d^2k_\perp}{16\pi^3} \no \\
&\hs{13mm}  \times             
\sum_{\l_q=+,-}\ \psi^{\da *}_{\l_q}(x,\vect{k}'_{\perp}) \psi^{\ua}_{\l_q}(x,\vect{k}_{\perp}),  \\[1.0ex]
\label{eq:fa2}
f^a_{2}(Q^2) &= \frac{2M}{q^1+i\,q^2}\int \frac{dx\,d^2k_\perp}{16\pi^3} \no \\
&\hs{0mm}  \times           
\sum_{\l_q,\,\l_a=+,-}\ \psi^{\da *}_{\l_q,\,\l_a}(x,\vect{k}'_{\perp}) \psi^{\ua}_{\l_q,\,\l_a}(x,\vect{k}_{\perp}).
\end{align}

Using the flavor wave function given in Eq.~\eqref{flave}, the quark flavor contributions to 
the bare (without pion cloud) proton Dirac form factor are therefore given by
\begin{align}
F^{(0),u}_{1p}(Q^2) &= \frac{3}{2}\,e_u\,f^s_1(Q^2) + \frac{1}{2}\,e_u\,f^a_1(Q^2), \\
F^{(0),d}_{1p}(Q^2) &= e_d\,f^a_1(Q^2),
\end{align}
where $e_u$ and $e_d$ are the quark charges and analogous expressions hold for the quark 
flavor contributions to the Pauli form factors, with $f^s_1 \to f^s_2$ and $f^a_1 \to f^a_2$.
The scalar functions
$f^s_1(Q^2)$ and $f^a_1(Q^2)$ are subject to the normalizations $f^s_{1}(0) = 1 = f^a_{1}(0)$, which 
guarantees the correct quark and hence nucleon charges. 
Using charge symmetry we obtain the following results for the bare nucleon Dirac form factors
\begin{align}
F_{1p}^{(0)}(Q^2) &= \frac{3}{2}\,e_u\,f^s_{1}(Q^2) + \frac{1}{2}\lf(e_u + 2\,e_d\rg) f^a_{1}(Q^2), \no \\
               &= f^s_{1}(Q^
               2),  \\
F_{1n}^{(0)}(Q^2) &= \frac{3}{2}\,e_d\,f^s_{1}(Q^2) + \frac{1}{2}\lf(e_d + 2\,e_u\rg) f^a_{1}(Q^2), \no \\
               &= -\frac{1}{2}\,f^s_1(Q^2) + \frac{1}{2}\, f^a_1(Q^2), 
\end{align}
where again analogous expressions hold for the Pauli form factors, with $f^s_1 \to f^s_2$ and $f^a_1 \to f^a_2$.

Using the LFWF given in Eq.~\eqref{eq:nucleon_LFWF2a} and the definition given in Eq.~\eqref{psis},
the explicit form of the scalar diquark helicity components of the LFWFs are
\begin{align}
\label{eq:lfwf_component1}
\sqrt{(1-x)}\,\psi^{\ua}_{+}(x,\vect{k}_\perp) &= \lf(M + \frac{m}{x}\rg)\varphi^s_1 + 2\,M\,\varphi^s_2, \\
\sqrt{(1-x)}\,\psi^{\ua}_{-}(x,\vect{k}_\perp) &= -\frac{1}{x}\lf(k^1 + i\,k^2\rg)\varphi^s_1, \\
\sqrt{(1-x)}\,\psi^{\da}_{+}(x,\vect{k}_\perp) &= \frac{1}{x}\lf(k^1 - i\,k^2\rg)\varphi^s_1, \\
\sqrt{(1-x)}\,\psi^{\da}_{-}(x,\vect{k}_\perp) &= \lf(M + \frac{m}{x}\rg)\varphi^s_1 + 2\,M\,\varphi^s_2.
\end{align}
Similarly, using Eq.~\eqref{eq:nucleon_LFWF2a} and the definition given in Eq.~\eqref{psiav}, 
the helicity components of the LFWFs for axial-vector diquark are
\begin{align}
\sqrt{(1-x)}\,\psi^{\ua}_{++}(x,\vect{k}_\perp) &= \frac{\sqrt{2}\lf(k^1 - i\,k^2\rg)}{x(1-x)}\varphi^a_1, \\
\sqrt{(1-x)}\,\psi^{\ua}_{-+}(x,\vect{k}_\perp) &= \sqrt{2}\lf(M + \frac{m}{x}\rg)\varphi^a_1 + 2\sqrt{2}\,M\,\varphi^a_2,\\
\sqrt{(1-x)}\,\psi^{\ua}_{+-}(x,\vect{k}_\perp) &= -\frac{\sqrt{2}\lf(k^1 + i\,k^2\rg)}{1-x}\varphi^a_1, \\
\sqrt{(1-x)}\,\psi^{\ua}_{--}(x,\vect{k}_\perp) &= 0,
\end{align}
and
\begin{align}
\sqrt{(1-x)}\,\psi^{\da}_{++}(x,\vect{k}_\perp) &= 0,  \\
\sqrt{(1-x)}\,\psi^{\da}_{-+}(x,\vect{k}_\perp) &= -\frac{\sqrt{2}\lf(k^1 - i\,k^2\rg)}{1-x}\varphi^a_1, \\
\sqrt{(1-x)}\,\psi^{\da}_{+-}(x,\vect{k}_\perp) &= -\sqrt{2}\lf(M + \frac{m}{x}\rg)\varphi^a_1 - 2\sqrt{2}\,M\,\varphi^a_2, \\
\label{eq:lfwf_component10}
\sqrt{(1-x)}\,\psi^{\da}_{--}(x,\vect{k}_\perp) &= \frac{\sqrt{2}\lf(k^1 + i\,k^2\rg)}{x(1-x)}\varphi^a_1.
\end{align}
Using these results it is then straightforward to obtain expressions for the scalar functions defined 
in Eqs.~\eqref{eq:fs1}-\eqref{eq:fa2}, namely
\begin{widetext}
\begin{align}
f^s_{1}(Q^2) &= \frac{1}{16\pi^3} \int \frac{dx\,d^2 k_\perp}{x^2(1-x)} \no \\
&\hs{4mm}\lf\{\lf[\vect{k}_\perp^2 + \lf(x\,M+m\rg)^2 - \frac{1}{4}(1-x)^2Q^2\rg]\vp^s_1\,\vp^s_1{}'
+ 2\,x\,M(x\,M + m)\lf(\varphi^s_1\,\varphi^s_2{}' + \varphi^s_1{}'\,\varphi^s_2\rg)
+ 4\,x^2\,M^2\,\varphi^s_2\,\varphi^s_2{}'\rg\}, \\[2.0ex]
f^a_{1}(Q^2) &= \frac{1}{8\pi^3} \int \frac{dx\,d^2 k_\perp}{x^2(1-x)}  \no \\
&\hs{-17mm} \biggl\{
\frac{1+x^2}{(1-x)^2}\lf[\vect{k}_\perp^2 - \frac{1}{4}(1-x)^2Q^2\rg]\vp^a_1\vp^a_1{}'
+ \lf(x\,M + m\rg)^2\vp^a_1\vp^a_1{}' 
+ 2\,x\,M\lf(x\,M+m\rg)\lf(\vp^a_1\vp^a_2{}' + \vp^a_1{}'\vp^a_2\rg) + 4\,x^2\,M^2\,\vp^a_2\vp^a_2{}' \biggr\}, \\[2.0ex]
f^s_{2}(Q^2) &= \frac{M}{8\pi^3} \int \frac{dx\,d^2 k_\perp}{x^2(1-x)} \no \\
&\hs{15mm}\lf[(1-x)(x\,M+m)\varphi^s_1{}'\,\varphi^s_1
-2x\,M\,\frac{\vect{k}_\perp\cdot\vect{q}_\perp }{Q^2}\lf(\varphi^s_1\,\varphi^s_2{}'-\varphi^s_2\,\varphi^s_1{}'\rg)
+ x(1-x)\,M\lf(\varphi^s_1\,\varphi^s_2{}'+\varphi^s_2\,\varphi^s_1{}'\rg)\rg], \displaybreak[2] \\
f^a_{2}(Q^2) &= -\frac{M}{4\pi^3} \int \frac{dx\,d^2 k_\perp}{x(1-x)^2} \no \\
&\hs{15mm}\lf[(1-x)(x\,M+m)\varphi^a_1{}'\,\varphi^a_1
-2x\,M\,\frac{\vect{k}_\perp\cdot\vect{q}_\perp }{Q^2}\lf(\varphi^a_1\,\varphi^a_2{}'-\varphi^a_2\,\varphi^a_1{}'\rg)
+ x(1-x)\,M\lf(\varphi^a_1\,\varphi^a_2{}'+\varphi^a_2\,\varphi^a_1{}'\rg)\rg],
\end{align}
\end{widetext}
where the prime refers to the final state wave functions. The invariant masses are then given by
\begin{multline}
M^2_0  = \frac{\lf(\vec{k}_\perp \mp \frac{1}{2}(1-x)\vec{q}_\perp\rg)^2 + m^2}{x} \\
       + \frac{\lf(\vec{k}_\perp \mp \frac{1}{2}(1-x)\vec{q}_\perp\rg)^2 + M_D^2}{1-x},
\end{multline}
where $M_D$ is the diquark mass, being either a scalar or axial-vector diquark, 
and the minus sign is for the initial state and plus sign the final state. 
Recall that $M$ is the nucleon mass and $m$ the constituent quark mass.

\subsection{Nucleon form factors with a pion cloud}
\begin{figure}[bp]
\centering\includegraphics[width=1.0\columnwidth,angle=0]{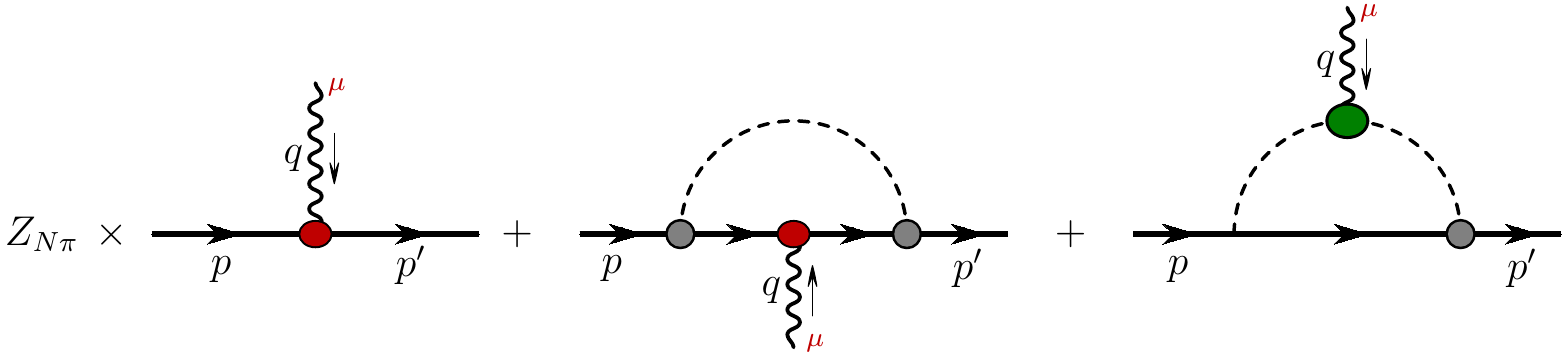}
\caption{Nucleon form factor diagrams, including the pion cloud. The multiplicative factor $Z_{N\pi}$
represents the probability that the nucleon is in a configuration without a pion cloud. In the second 
diagram the photon couples to the bare nucleon and in the third diagram it couples to the pion.}
\label{fig:pion_nucleon}
\end{figure}

The pion cloud component of our model for the nucleon is introduced via a single
pion loop around our bare nucleon, as illustrated in Fig.~\ref{fig:pion_nucleon} for
the nucleon electromagnetic current. The first diagram represents the photon coupling to the
bare nucleon, multiplied by $Z_{N\pi}$, 
which represents the probability that the nucleon 
is in a configuration without a pion cloud.
The second diagram in Fig.~\ref{fig:pion_nucleon}
represents the photon coupling to the bare nucleon with a pion in the air, the photon
coupling is given by
\begin{align}
\L^\mu(p',p) &= \tfrac{1}{2}\lf(1+\tau_3\rg)\lf[\g^\mu\,F^{(0)}_{1p}(Q^2) + \frac{i\s^{\mu\nu}q_\nu}{2\,M}\,F^{(0)}_{2p}(Q^2)\rg] \no \\
&\hs{-9mm}
+ \tfrac{1}{2}\lf(1-\tau_3\rg)\lf[\g^\mu\,F^{(0)}_{1n}(Q^2) + \frac{i\s^{\mu\nu}q_\nu}{2\,M}\,F^{(0)}_{2n}(Q^2)\rg],
\end{align}
where $F^{(0)}_{1p}(Q^2)$, $F^{(0)}_{2p}(Q^2)$, etc,  are the bare nucleon form factors 
discussed in Sect.~\ref{sec:bare}. 
The contribution
of this second diagram to observable quantities is usually small. Finally, the third 
diagram in Fig.~\ref{fig:pion_nucleon} represents the photon coupling to the pion in the loop,
with a pion electromagnetic vertex given by
\begin{align}
\L^\mu_{ij}(p',p) = \ve_{3ji}\,\lf(p' + p\rg)^\mu\,F_\pi(Q^2),
\end{align}
where the pion form factor has the form $F_\pi(Q^2) = \lf[1+Q^2/\L_\pi^2\rg]^{-1}$ and we choose 
the standard value of $\L_\pi^2 = 0.5\,$GeV$^2$.

The complete expressions for the proton and neutron Dirac and Pauli form factors are then
\begin{align}
\label{fpc}
F_{1p} &= Z_{N\pi}\,F^{(0)}_{1p}
+ \lf(\tfrac{1}{2}\,F^{(0)}_{1p} + F^{(0)}_{1n}\rg) F^{(N),\text{vec}}_{1N} \no \\
&\hs{8mm}
+ \lf(\tfrac{1}{2}\,F^{(0)}_{2p} + F^{(0)}_{2n}\rg) F^{(N),\text{ten}}_{1N} + F^{(\pi)}_{1N}, \displaybreak[3] \\[1.0ex] 
F_{1n}(Q^2) &= Z_{N\pi}\,F^{(0)}_{1n} + \lf(F^{(0)}_{1p} + \tfrac{1}{2}\,F^{(0)}_{1n}\rg) F^{(N),\text{vec}}_{1N} \no \\
&\hs{1mm}
+ \lf(F^{(0)}_{2p} + \tfrac{1}{2}\,F^{(0)}_{2n}\rg) F^{(N),\text{ten}}_{1N}(Q^2) - F^{(\pi)}_{1N},  \\[1.0ex]
F_{2p}(Q^2) &= Z_{N\pi}\,F^{(0)}_{2p} + \lf(\tfrac{1}{2}\,F^{(0)}_{1p} + F^{(0)}_{1n}\rg) F^{(N),\text{vec}}_{2N} \no \\
&\hs{7mm}
+ \lf(\tfrac{1}{2}\,F^{(0)}_{2p} + F^{(0)}_{2n}\rg) F^{(N),\text{ten}}_{2N} + F^{(\pi)}_{2N}, \\[1.0ex]
\label{fpc1}
F_{2n}(Q^2) &= Z_{N\pi}\,F^{(0)}_{2n} 
+ \lf(F^{(0)}_{1p} + \tfrac{1}{2}\,F^{(0)}_{1n}\rg) F^{(N),\text{vec}}_{2N} \no \\
&\hs{7mm}
+ \lf(F^{(0)}_{2p} + \tfrac{1}{2}\,F^{(0)}_{2n}\rg) F^{(N),\text{ten}}_{2N}
- F^{(\pi)}_{2N},
\end{align}
where the $Q^2$ dependence of the various form factors has been omitted for clarity.
The form factors $F^{(N),\text{vec}}_{1N}(Q^2)$ and $F^{(N),\text{vec}}_{2N}(Q^2)$ result from the second diagram
in Fig.~\ref{fig:pion_nucleon} where the photon couples to the bare nucleon with a $\gamma^\mu$, while
the form factors $F^{(N),\text{ten}}_{1N}(Q^2)$ and $F^{(N),\text{ten}}_{2N}(Q^2)$ arise from the $i\s^{\mu\nu}q_\nu$
coupling to the bare nucleon in the same diagram. Recall $F^{(0)}_{1p}$, $F^{(0)}_{1n}$, etc, are the bare
nucleon form factors discussed in Sect.~\ref{sec:bare}. 

For the form factors arising from the pion loop in the second diagram 
of Fig.~\ref{fig:pion_nucleon} we find~\cite{Matevosyan:2005bp}
\begin{align}
&F^{(N),\text{vec}}_{1N}(Q^2) = g_{\pi N}^2 \,
\int_0^1 dx\, \int \frac{d^2k_\perp}{2(2\pi)^3}  \no \\
&\hs{3mm}
F_{\pi N}(\vect{\ell}_+^2,x)\,F_{\pi N}(\vect{\ell}_-^2,x) \
\frac{x\lf[\vect{k}_\perp^2 + x^2\,M^2 - \tfrac{1}{4}\,x^2\,Q^2\rg]}
{D^+(\vect{k}_\perp)\ D^-(\vect{k}_\perp)}, \\
&F^{(N),\text{vec}}_{2N}(Q^2) = -2\,g_{\pi N}^2\,M^2
\int_0^1 dx \int \frac{d^2k_\perp}{2(2\pi)^3}  \no \\ 
&\hs{10mm}
F_{\pi N}(\vect{\ell}_+^2,x)\,F_{\pi N}(\vect{\ell}_-^2,x) \
\frac{x^3}{D^+(\vect{k}_\perp)\ D^-(\vect{k}_\perp)},
\end{align}
and 
\begin{align}
&F^{(N),\text{ten}}_{1N}(Q^2) = -g_{\pi N}^2\,\frac{1}{2} \,
\int_0^1 dx \int \frac{d^2k_\perp}{2(2\pi)^3} \no \\
&\hs{15mm}
F_{\pi N}(\vect{\ell}_+^2,x)\,F_{\pi N}(\vect{\ell}_-^2,x) \
\frac{x^3\,Q^2} {D^+(\vect{k}_\perp)\ D^-(\vect{k}_\perp)}, \\[1.0ex]
&F^{(N),\text{ten}}_{2N}(Q^2) = -g_{\pi N}^2 \,
\int_0^1 dx \int \frac{d^2k_\perp}{2(2\pi)^3} \no \\
&\hs{0mm} 
F_{\pi N}(\vect{\ell}_+^2,x)\,F_{\pi N}(\vect{\ell}_-^2,x) \
\frac{x\lf[x^2\,M^2 - \tfrac{1}{4}\,x^2\,Q^2 + k_x^2 - k_y^2\rg]}
{D^+(\vect{k}_\perp)\ D^-(\vect{k}_\perp)}.
\end{align}
where we take $g_{\pi N} = 13.5$, $\vect{\ell}_\pm \equiv \vect{k}_\perp \pm \frac{1}{2}\,x\,\vect{q}_\perp$ and
\begin{align}
D^{\pm}(\vect{k}_\perp) &= \lf(\vect{k}_\perp \pm \tfrac{1}{2}\,x\,\vect{q}_\perp\rg)^2 + x^2\,M^2  + \lf(1-x\rg)m_\pi^2,
\end{align}
with $m_\pi$ the pion mass.
The pion-nucleon form factor that enters diagrams two and three in 
Fig.~\ref{fig:pion_nucleon} is taken to be
\begin{align}
F_{\pi N}\lf(\ell_\perp^2,x\rg) = e^{-\lf[\ell_\perp^2 + x^2\,M^2 + \lf(1-x\rg)m_\pi^2\rg]/\lf[2x\lf(1-x\rg)\L^2\rg]},
\label{pin}
\end{align}
where $\L$ is a parameter that encapsulates the non-pointlike nature of the
pion-nucleon vertex and will be determined in Sect.~\ref{sec:res}.
The form of $ F_{\pi N}$ is chosen so as to maintain charge and momentum conservation~\cite{Szczurek:1996ur}.
An improvement was suggested in~\cite{Alberg:2012wr}, but this has not yet been applied to calculating 
electromagnetic form factors.

The form factors arising from the pion loop in the third diagram 
of Fig.~\ref{fig:pion_nucleon} are given by~\cite{Matevosyan:2005bp}
\begin{align}
&F^{(\pi)}_{1N} = g_{\pi N}^2\,F_\pi(Q^2)\ \int_0^1 dx \int \frac{d^2k_\perp}{(2\pi)^3}  \no \\
&
F_{\pi N}(\vect{k}_+^2,x)\,F_{\pi N}(\vect{k}_-^2,x) 
\frac{x\lf[
\vect{k}^2_\perp  - \tfrac{1}{4}(1-x)^2Q^2 + x^2 M^2 \rg]}
{D^{+}(\vect{k}_\perp) \ D^{-}(\vect{k}_\perp)}, \\[1.0ex]
&F^{(\pi)}_{2N} = 2\,g_{\pi N}^2\,M^2\,F_\pi(Q^2)\ \int_0^1 dx \int \frac{d^2k_\perp}{(2\pi)^3} \no \\
&\hs{12mm}
F_{\pi N}(\vect{k}_+^2,x)\,F_{\pi N}(\vect{k}_-^2,x)
\frac{x^2\lf(1-x\rg)}{D^{+}(\vect{k}_\perp) \ D^{-}(\vect{k}_\perp)},
\end{align}
where we have defined
\begin{align}
\vect{k}_\pm &\equiv \vect{k}_\perp \pm \frac{1}{2}\,(1-x)\,\vect{q}_\perp.
\end{align}

We note that the present version provides a minimal treatment of the pion cloud. Effects of the intermediate 
$\Delta$ and   terms involving a $\gamma N\to\pi N$ direct coupling  are not included. Both of these terms 
involve distances smaller than those of the terms we do include, which dominate in the chiral limit. 
Therefore we shall assume that such effects are subsumed within the parameters of the model. We shall 
see that achieving the present modest goal of 
reproducing form factors, while remaining consistent with the small fraction of the nucleon total angular 
momentum carried by the quarks spin is possible without including terms  additional to those of the above equations.

\section{Results For Nucleon form factors and their flavor dependence}
\label{sec:res}

\begin{table*}[tbp]
\begin{center}
\addtolength{\tabcolsep}{5.0pt}
\addtolength{\extrarowheight}{1.8pt}
\begin{tabular}{c|cccccccccc|ccc}
\hline\hline
$\chi^2$   & $m$   & $M_s$  & $M_a$   & $c_s$ & $\beta_s$ & $\g_s$  & $c_a$ & $\beta_a$ & $\g_a$  & $\L$   & $\mu_p~(\mu_N)$ & $\mu_n~(\mu_N)$\\
\hline
0.078516   & 0.191 & 0.414 & 0.167 & 1.509 & 1.226     & 5.719  & 0.008 & 1.104     & 8.586  & 1.035 & 2.794   & -1.849 \\
\hline\hline
\end{tabular}
\caption{Model parameters: $m$ constituent quark mass, $M_s$ scalar diquark mass, $M_a$ axial vector diquark mass,
quark--scalar-diquark nucleon LFWF parameters $c_s$,  $\beta_s$ , $\g_s$ (see \eq{phi}),
quark--axial-vector-diquark nucleon LFWF parameters $c_a$,  $\beta_a$ , $\g_a$ (see \eq{phi}),
pion-nucleon vertex parameter $\Lambda$ (see \eq{pin}). All mass-dimensioned parameters are in GeV. The first column gives
the $\chi^2$ obtained in the fit expressed in Eq.~\eqref{eq:chisquared} and the final two columns present our results for the proton
and neutron magnetic moments.} 
\end{center}
\label{tab:parameters}
\end{table*}

\begin{figure}[tbp]
\centering\includegraphics[width=1.0\columnwidth,angle=0]{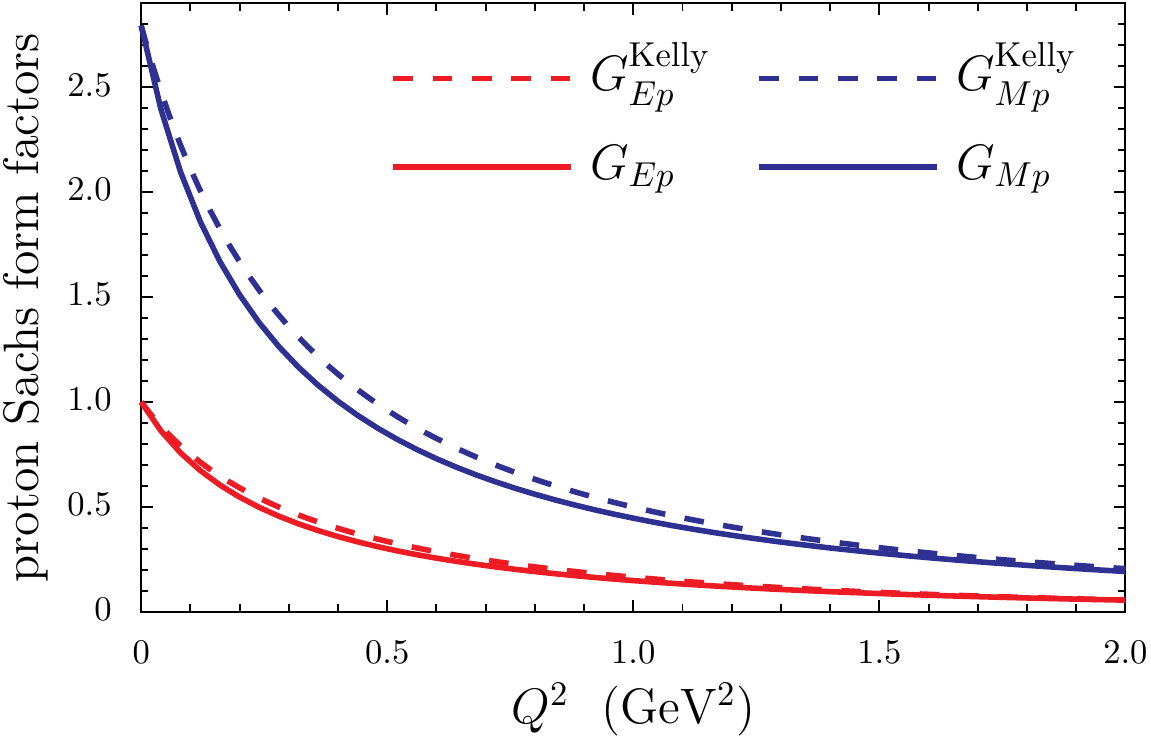}
\caption{(Color online) Solid lines are the model results for the proton Sachs form factors and the dashed
lines are the empirical results from Kelly given in Ref.~\cite{Kelly:2004hm}.}
\label{fig:protonsachs}
\end{figure}

\begin{figure}[tbp]
\centering\includegraphics[width=1.0\columnwidth,angle=0]{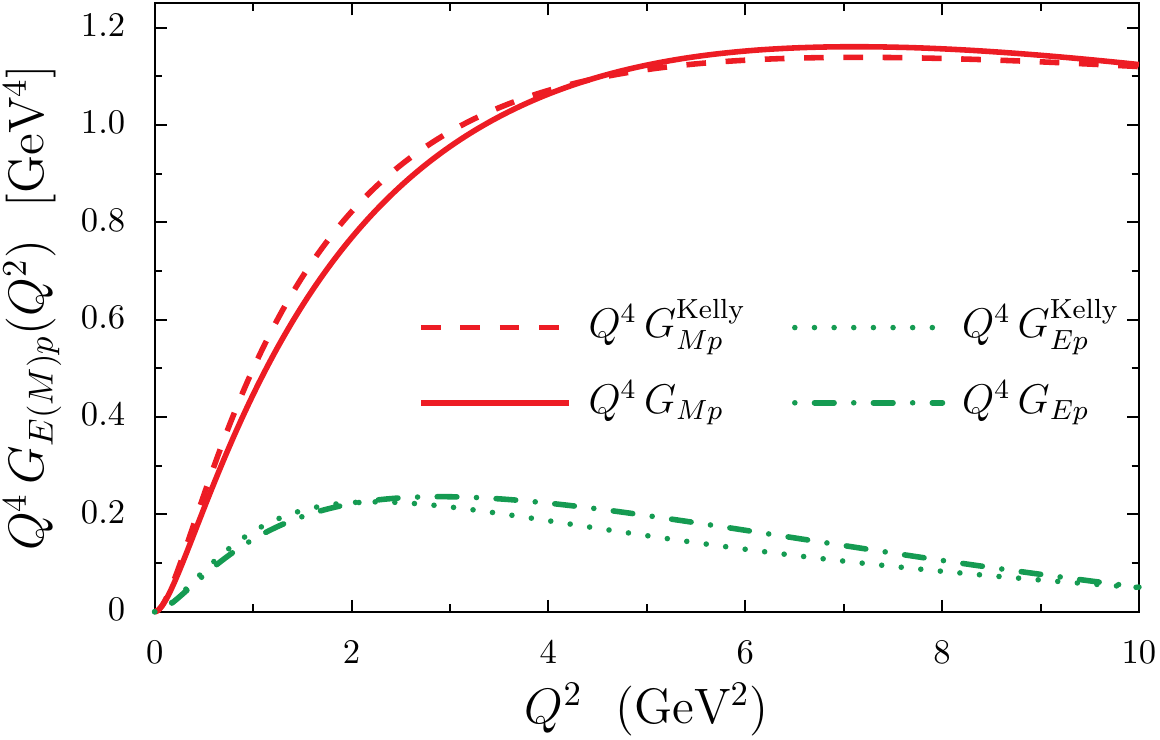}
\caption{(Color online) Proton Sachs form factors and their comparison with the empirical parametrizations
of Ref.~\cite{Kelly:2004hm}.}
\label{fig:GMhigh}
\end{figure}

\begin{figure}[tbp]
\centering\includegraphics[width=1.0\columnwidth,angle=0]{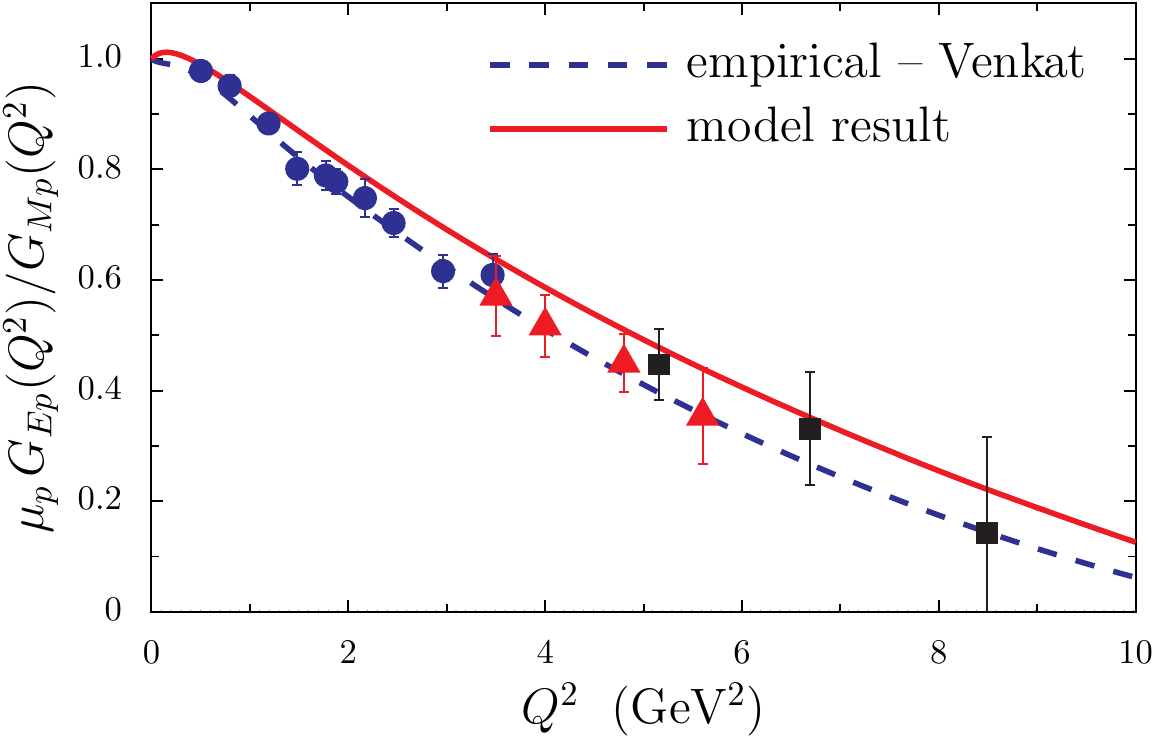}
\caption{(Color online) Ratios of the proton electric to magnetic Sachs form factors. The solid curve is our model result
and the dashed curve is the phenomenological fit of Ref.~\cite{Venkat:2010by}.
The data are from Refs.~\cite{Jones:1999rz,Gayou:2001qt,Gayou:2001qd,Puckett:2010ac,Puckett:2011xg}.}
\label{fig:GEGMproton}
\end{figure}

\begin{figure}[tbp]
\centering\includegraphics[width=1.0\columnwidth,angle=0]{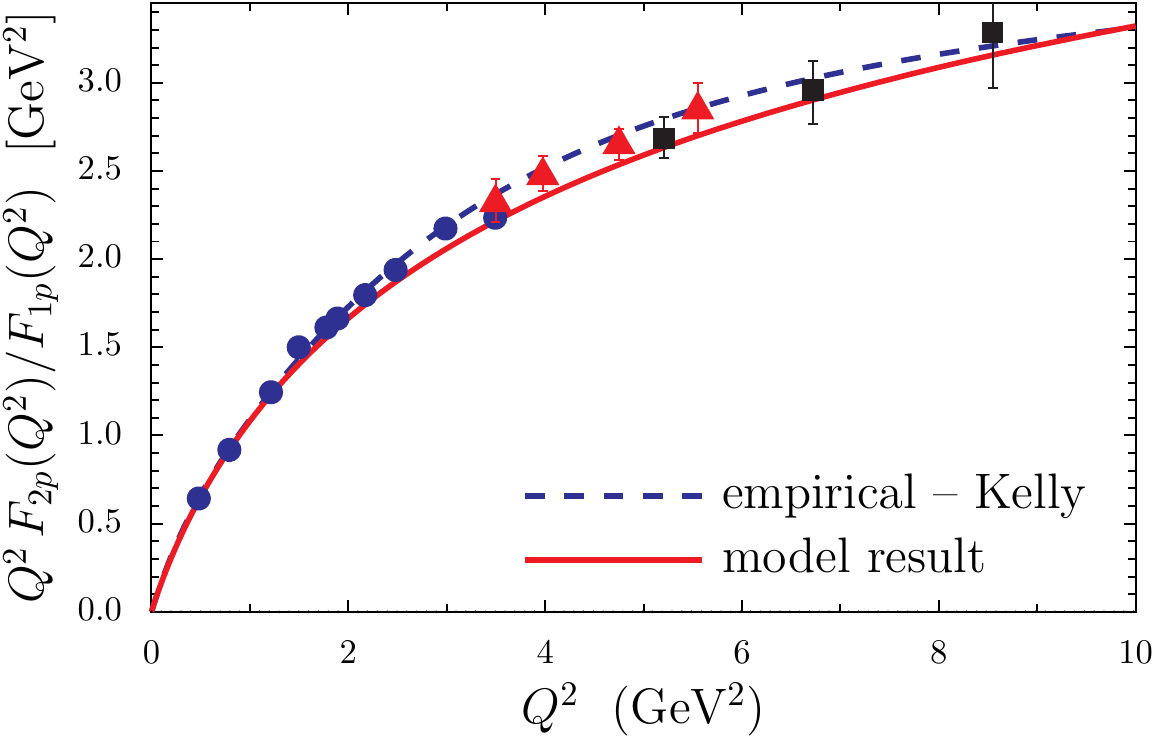}
\caption{(Color online) Ratios of the proton Pauli to Dirac form factors multiplied by $Q^2$. The solid curve is our model result
and the dashed curve is the empirical result of Ref.~\cite{Kelly:2004hm}.
The data are from Refs.~\cite{Jones:1999rz,Gayou:2001qt,Gayou:2001qd,Puckett:2010ac,Puckett:2011xg}.}
\label{fig:F1F2proton}
\end{figure}

The parameters of the model are as follows: the quark, scalar diquark and axial-vector diquark masses, labeled
by $m$, $M_s$ and $M_a$ respectively; the three parameters $c_s$, $\beta_s$ and $\g_s$ (see Eq.~\eqref{phi}) that 
specify the quark--scalar-diquark component of the nucleons LFWF and the analogous three parameters 
$c_a$, $\beta_a$ and $\g_a$ which encapsulates the quark--axial-vector-diquark component of the nucleons LFWF.
Finally, there is the parameter $\L$ which enters Eq.~\eqref{pin} and describes the high momentum transfer fall 
off of the pion-nucleon vertex function. Therefore, in total the model has ten parameters and
these are chosen to minimize $\chi^2$ as defined by
\begin{multline}
\chi^2 \equiv \frac{1}{4\,n_q}
\sum_{Q^2} \Biggl[ \frac{|F_{1p} - F_{1p}^{\text{exp}}|}{|F_{1p}^{\text{exp}}|} + \frac{|F_{2p}-F_{2p}^{\text{exp}}|}{|F_{2p}^{\text{exp}}|} \\
+ \frac{|F_{1n} - F_{1n}^{\text{exp}}|}{|F_{1n}^{\text{exp}}|} + \frac{|F_{2n}-F_{2n}^{\text{exp}}|}{|F_{2n}^{\text{exp}}|} \Bigr],
\label{eq:chisquared}
\end{multline}
where $F_{1p}$, etc, are the form factors from the model, given in Eqs.~\eqref{fpc}-\eqref{fpc1}, and for 
the empirical form factors, namely $F_{1p}^{\text{exp}}$, etc,
we take the results from Ref.~\cite{Kelly:2004hm}. For the sum in Eq.~\eqref{eq:chisquared} we take
$n_q$ values of $Q^2$ chosen uniformly on the domain $Q^2 \in [0,10]\,$GeV$^2$ and 
Table~\ref{tab:parameters} gives the resulting model parameters for $n_q=11$. 
The resulting values of the nucleon magnetic moments are also shown in the table and are in good agreement with 
the experimental values of $\mu_p = 2.79\,\mu_N$ and $\mu_n = -1.91\,\mu_N$ for the proton and neutron, respectively.

\begin{figure}[tbp]
\centering\includegraphics[width=1.0\columnwidth,angle=0]{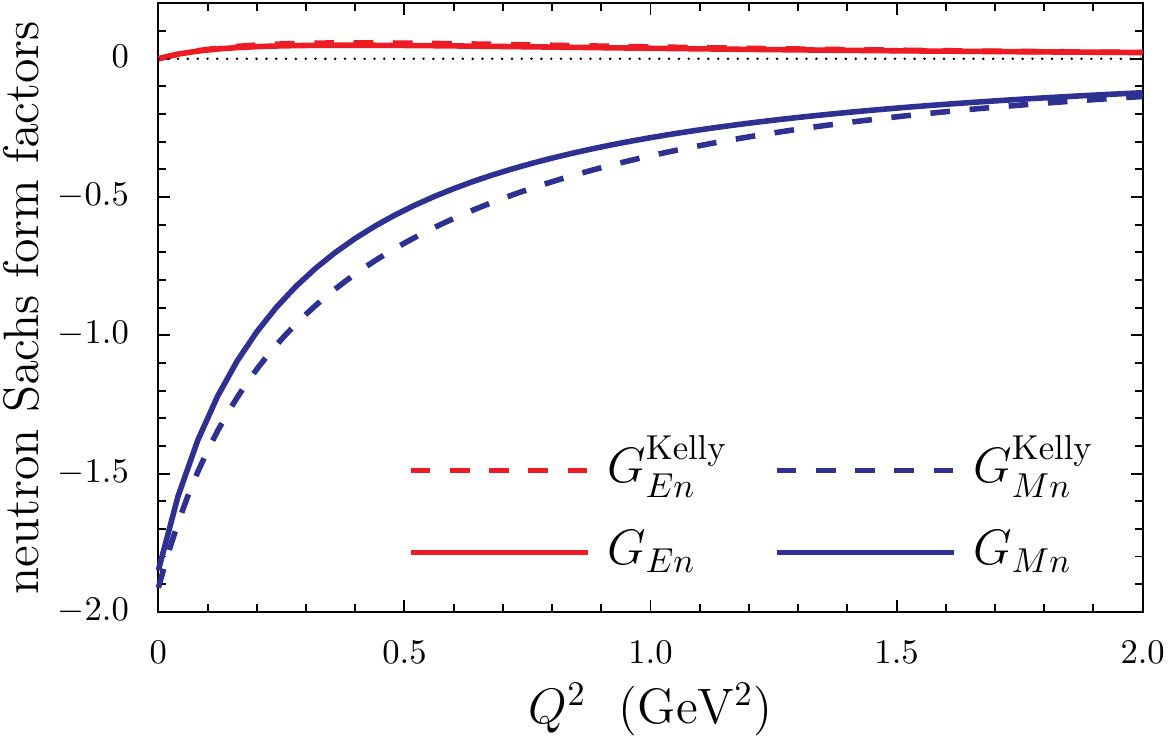}
\caption{(Color online) Solid lines are the model results for the neutron Sachs form factors and the dashed
lines are the empirical results from Kelly given in Ref.~\cite{Kelly:2004hm}.}
\label{fig:neutronsachs}
\end{figure}

\begin{figure}[tbp]
\centering\includegraphics[width=1.0\columnwidth,angle=0]{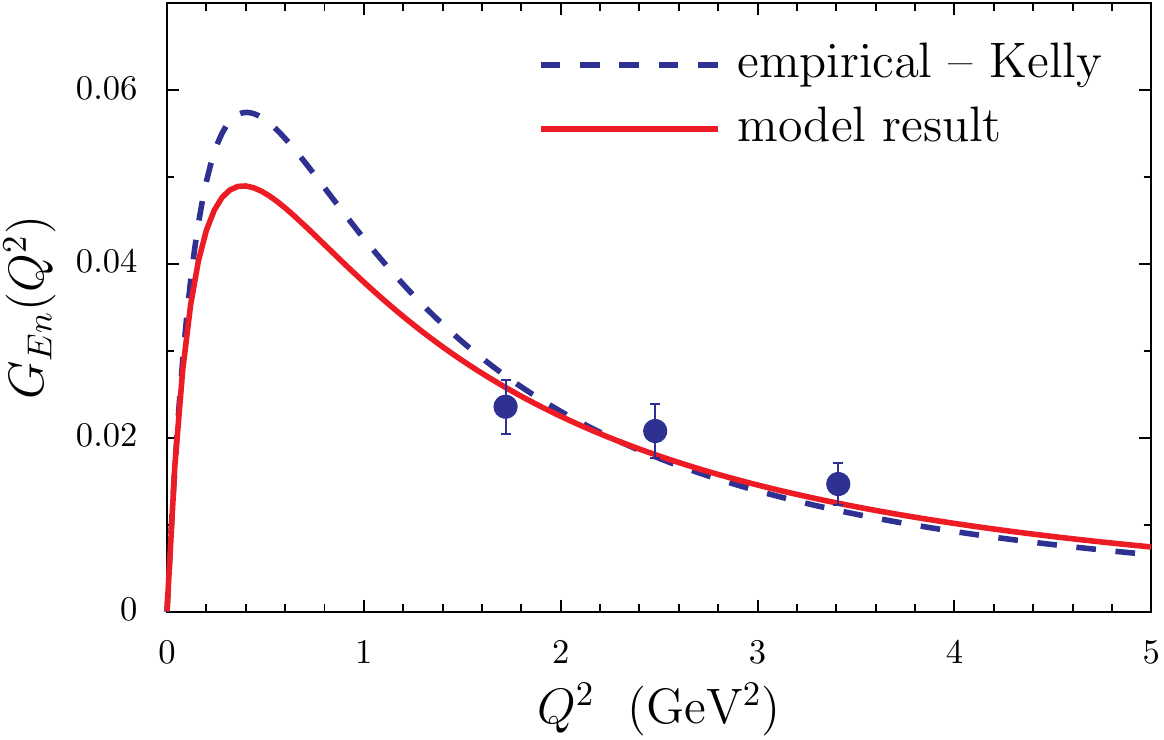}
\caption{(Color online) The model result for the neutron Sachs electric form factor is given by the solid
line and the dashed curves is from Kelly~\cite{Kelly:2004hm}.
The data is from Ref.~\cite{Riordan:2010id}.}
\label{fig:gen}
\end{figure}

The results for the proton Sachs form factors are shown in Figs.~\ref{fig:protonsachs} and \ref{fig:GMhigh}. We find 
that our results agree very well, over a large $Q^2$ range, with the empirical parameterizations of Kelly 
given in Ref.~\cite{Kelly:2004hm}. At small $Q^2$ both the electric and magnetic form factors fall
off a little too rapidly, which is a likely indication that the pion cloud component of the LFWF is slightly too
large. In Figs.~\ref{fig:GEGMproton} and \ref{fig:F1F2proton} we compare our form factor results with data for 
the ratios $\mu_p\,G_{Ep}(Q^2)/G_{Mp}(Q^2)$ and $Q^2\,F_{2p}(Q^2)/F_{1p}(Q^2)$, respectively. In each case our results
agree very well with the measured ratios, and although not shown in Fig.~\ref{fig:GEGMproton} we find that
the $G_{Ep}/G_{Mp}$ form factor ratio crosses zero at $Q^2 \simeq 12.3\,$GeV$^2$.

\begin{figure}[tbp]
\centering\includegraphics[width=1.0\columnwidth,angle=0]{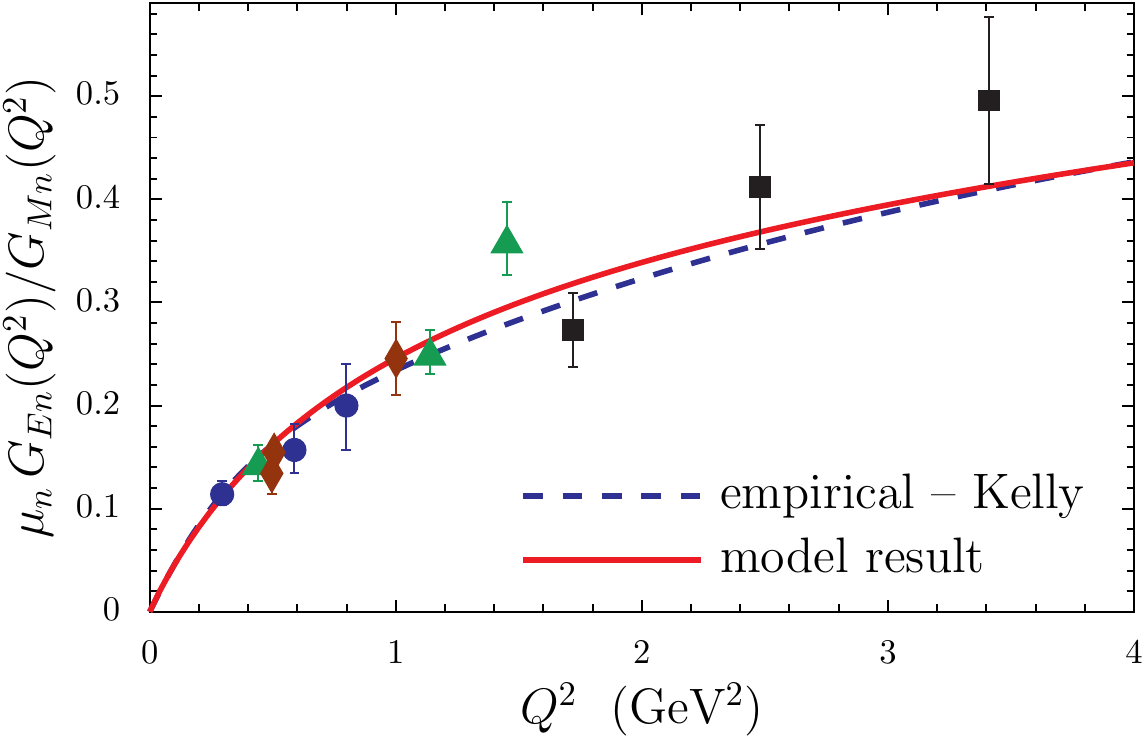}
\caption{(Color online) Ratios of the neutron electric to magnetic Sachs form factors. The solid curve is our model result
and the dashed curve is the phenomenological fit of Ref.~\cite{Kelly:2004hm}.
The data are from Ref.~\cite{Riordan:2010id}.}
\label{fig:GEGMneutron}
\end{figure}

\begin{figure}[tbp]
\centering\includegraphics[width=1.0\columnwidth,angle=0]{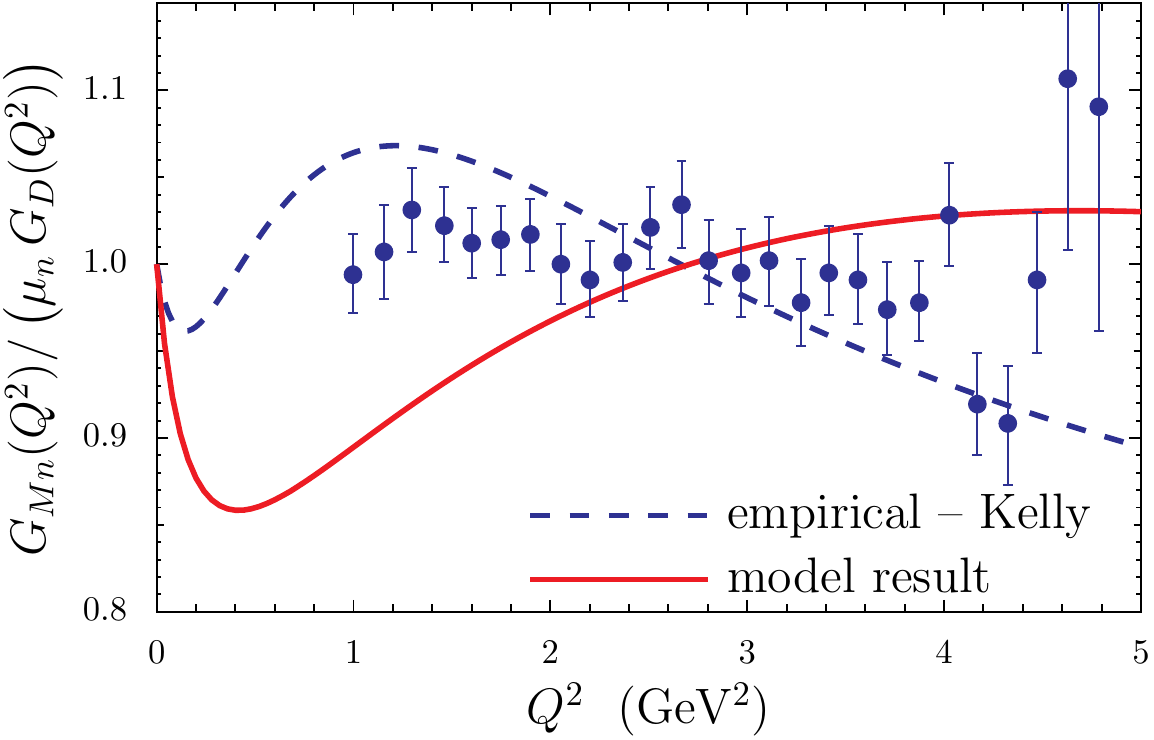}
\caption{(Color online) The solid curve is our model result for the neutron form factors ratio of
$G_{Mn}/\lf(\mu_n\, G_D\rg)$, where $G_D(Q^2)$ is the dipole form factor with mass parameter $\L = 0.71\,$GeV$^2$.
The dashed curve is the empirical result from Ref.~\cite{Kelly:2004hm} and the data is from Ref.~\cite{Lachniet:2008qf}.}
\label{fig:gmn}
\end{figure}

Our results for the neutron Sachs form factors are illustrated in Figs.~\ref{fig:neutronsachs} and \ref{fig:gen}.
A comparison with the empirical parameterizations of Ref.~\cite{Kelly:2004hm} and the recent Jefferson Lab
data for $G_E^n$, given in Ref.~\cite{Riordan:2010id}, shows excellent agreement. Similar to the proton case,
we find that our neutron magnetic form factor falls slightly too fast for small values of $Q^2$. However, our agreement
with the Kelly result for $G_{En}$ is extremely good.
Figs.~\ref{fig:GEGMneutron} and \ref{fig:gmn} compare our form factor results with data for 
the ratios $\mu_n\,G_{En}(Q^2)/G_{Mn}(Q^2)$ and $G_{Mn}(Q^2)/\lf[\mu_n\,G_{D}(Q^2)\rg]$, respectively,
where $G_D(Q^2)$ is the dipole form factor with mass parameter $\L = 0.71\,$GeV$^2$.
The comparison between data and our model results in Fig.~\ref{fig:GEGMneutron} is very good and our 
description of the $G_{Mn}(Q^2)$ data from Ref.~\cite{Lachniet:2008qf} (see Fig.~\ref{fig:gmn}) 
is generally as good as the one provided by Kelly~\cite{Kelly:2004hm} and seems to be better 
for the larger values of $Q^2$. 

\begin{figure}[tbp]
\centering\includegraphics[width=1.0\columnwidth,angle=0]{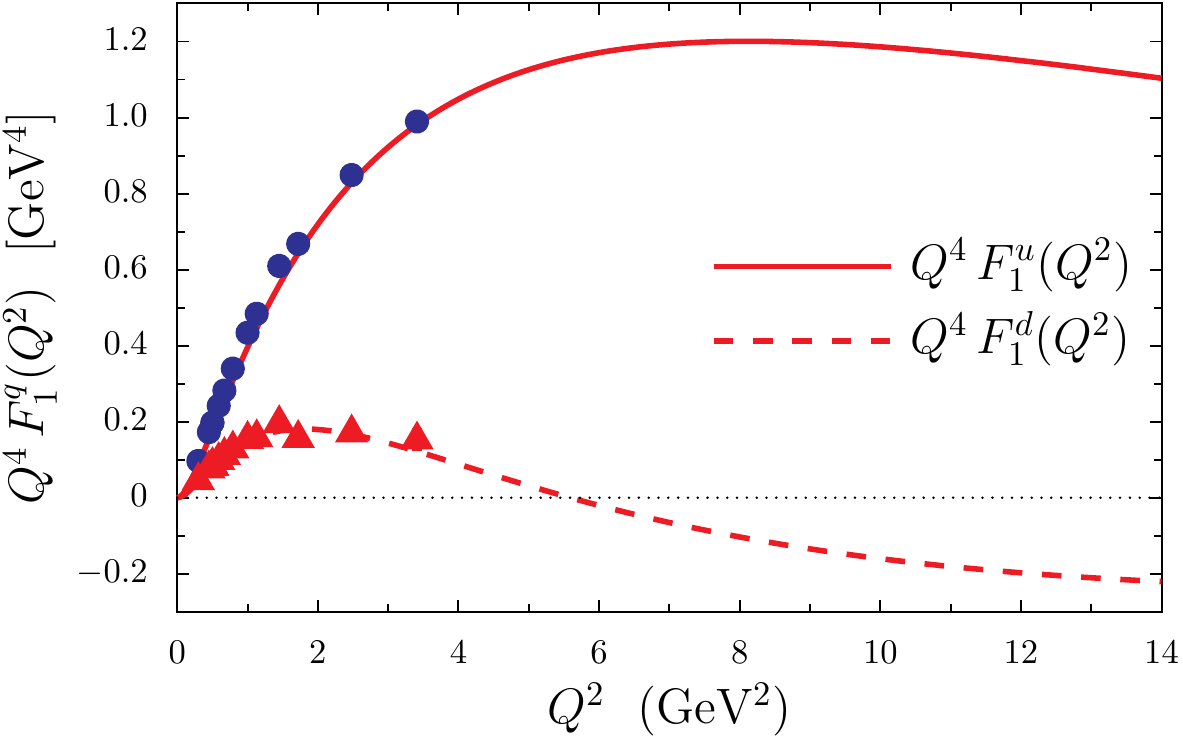}
\caption{(Color online) Model results for the Dirac quark sector form factors $F_1^u$ and $F_1^d$ multiplied by 
$Q^4$. The  data are 
from~\cite{Cates:2011pz,Jones:1999rz,Gayou:2001qt,Gayou:2001qd,Puckett:2010ac,Puckett:2011xg,Riordan:2010id,
Zhu:2001md,Bermuth:2003qh,Warren:2003ma,Glazier:2004ny,Plaster:2005cx} .}
\label{fig:f1q}
\end{figure}
\begin{figure}[tbp]
\centering\includegraphics[width=1.0\columnwidth,angle=0]{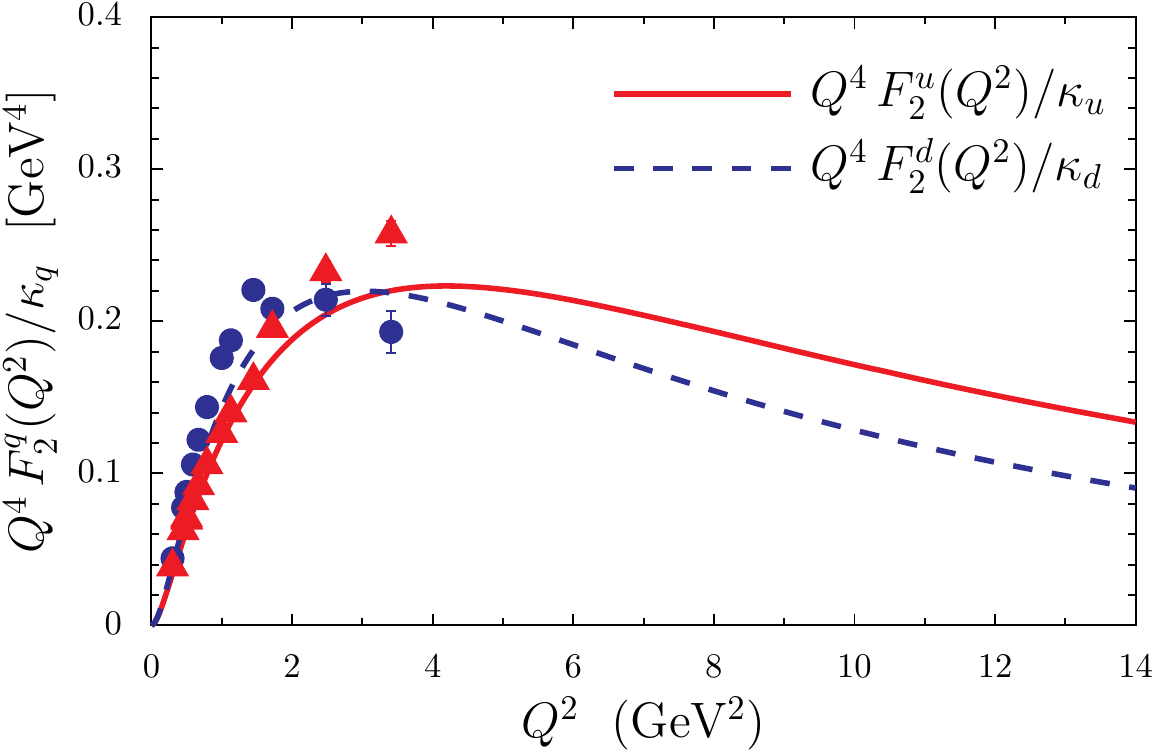}
\caption{(Color online) Model results for the Pauli quark sector form factors $F_2^u$ and $F_2^d$ multiplied by 
$Q^4$. The data are 
from~\cite{Cates:2011pz,Jones:1999rz,Gayou:2001qt,Gayou:2001qd,Puckett:2010ac,Puckett:2011xg,
Riordan:2010id,Zhu:2001md,Bermuth:2003qh,Warren:2003ma,Glazier:2004ny,Plaster:2005cx} .}
\label{fig:f2q}
\end{figure}

The importance of looking at the separate quark sector form factors for $u$ and $d$ quarks in the nucleon has been stressed in
Ref.~\cite{Cates:2011pz}. This is possible because of the charge symmetry (invariance under interchange of $u$ and $d$ quarks) 
of the nucleon wave function~\cite{Miller:1990iz,Miller:2006tv}. The quark sector Dirac and Pauli form factors are defined by
\begin{align}
F^u_{1(2)} = 2\,F^u_{1(2)p} + F^u_{1(2)n} \quad \text{\&} \quad F^d_{1(2)} = F^d_{1(2)p} + 2\,F^d_{1(2)p}.
\end{align}
We illustrate results for $Q^4\,F_1^q$ and $Q^4\,F_2^q/\kappa_q$, where $\kappa_q \equiv F^q_{2}(Q^2=0)$ and $q \in u,\,d$,
in Figs.~\ref{fig:f1q} and \ref{fig:f2q}, respectively. In each case the agreement between our results and the data 
from Ref.~\cite{Cates:2011pz} is very good. We predict that $F_1^d$ has a zero-crossing at approximately $5.5\,$GeV$^2$
and also observe a cross over between $F_2^u$ and $F_2^d$ at approximately $3\,$GeV$^2$.
This is consistent with the data in Ref.~\cite{Cates:2011pz}, where it is shown that for both the 
Dirac and Pauli quark sector form factors, the $d$ quark sector drops faster than the $u$ quark sector.
The data in Ref.~\cite{Cates:2011pz} also exhibits the behavior that on the domain $1\,$GeV$^2 \lesssim Q^2 \lesssim 3.4\,$GeV$^2$, the ratio of the Pauli 
to Dirac form factors, in both the $u$- and $d$-quark sectors, is almost constant.

\section{Proton spin content \label{sec:spin}}
The true test of this model is the independent prediction of the proton spin content. This prediction is implied 
by the definition of 
the flavor-spin wave function given in  \eq{flave} and the LFWFs given in Eqs.~\eqref{psis} and \eqref{psiav}.
The helicity parton distribution functions (PDFs) are given by
\begin{align}
\Delta q(x) = q_+(x) - q_-(x),
\label{eq:pdfhelicity}
\end{align}
where $q_+(x)$ is the number density of quarks with helicity parallel to the nucleon spin and $q_-(x)$ is the 
number density of quarks with helicity anti-parallel to the nucleon spin. The quark spin content, namely
$\Delta \Sigma = \Delta u + \Delta d$, is obtained 
by integrating Eq.~\eqref{eq:pdfhelicity} over $x$ for both the $u$ and $d$ quarks. In this work we ignore
contributions to $\Delta \Sigma$ from the heavier quark flavors.

Using the proton spin-flavor wave function of \eq{flave}, we obtain
\begin{align}
\Delta u(x) &= \frac{3}{2}\,\Delta q_s(x) + \frac{1}{2}\,\Delta q_a(x), \\
\Delta d(x) &= \Delta q_a(x),  
\end{align}
for the bare nucleon, 
where the subscripts $s$ and $a$ refer to the contributions to the helicity PDFs from the 
quark--scalar-diquark and quark--axial-vector-diquark components of the nucleons LFWF. The functions
$\Delta q_s(x)$ and $\Delta q_a(x)$ are completely analogous to the bare nucleon form factor
quantities $f^s_1(Q^2)$ and $f^a_1(Q^2)$, respectively, and expressions can easily by obtained using Eq.~\eqref{eq:pdfhelicity} and
the results given in Eqs.~\eqref{eq:lfwf_component1}-\eqref{eq:lfwf_component10}. We find
\begin{align}
\label{eq:deltaqs}
\Delta q_s(x) &= \frac{Z_s}{16\pi^3} \int \frac{d^2k_\perp}{x^2(1-x)} \no \\
&\hs{4mm}
\left[\lf[\lf(M\,x+m\rg)\vp^s_1+2\,M\,x\,\vp^s_2\right]^2 - k_\perp^2\vp^s_2{}^2\right], \\
\label{eq:deltaqa}
\Delta q_a(x) &= \frac{Z_a}{8\pi^3}  \int \frac{d^2k_\perp}{x^2(1-x)} \no \\
&\hs{-8mm}
\left[\frac{1+x^2}{(1-x)^2}k_\perp^2 \vp^a_1{}^2 - \left[(M\,x+m)\vp^a_1+2\,M\,x\,\vp^a_2\right]^2\right].
\end{align}

The spin content is determined by the first moments of the helicity PDFs, namely
\begin{align}
\Delta u &\equiv \int_0^1 dx\, \Delta u(x) = \frac{3}{2} \Delta q_s + \frac{1}{2}\Delta q_a, \\
\Delta d &\equiv \int_0^1 dx\, \Delta d(x) = \Delta q_a.
\end{align}
The dominant terms in Eqs.~\eqref{eq:deltaqs} and \eqref{eq:deltaqa} are those containing the nucleon mass, 
and these come in with a positive sign for the scalar diquark component and with a negative sign for the axial-vector piece. 
This implies that the quark spin content of the term with the axial-vector diquark can be expected to be negative. 
Importantly, these results refer to the contribution of the nucleon without including the effects of 
the pion cloud.

The effect of the pion cloud on the nucleon spin sum is determined by evaluating the diagrams illustrated 
in Fig.~\ref{fig:pion_nucleon}, where instead of the electromagnetic current operator we insert the quark 
spin operator. In this case, only the first and second diagrams in Fig.~\ref{fig:pion_nucleon} contribute
because the spin of the pion zero. We find that the nucleon spin sum, including the pion cloud, is given
by
\begin{align}
\Delta \Sigma_\pi = \lf(Z_{N \pi} + \Delta q^\pi_N\rg)\lf(\Delta u +\Delta d\rg),
\label{all}
\end{align}
where
\begin{multline}
\Delta q^\pi_N = -3\, g_{\pi N}^2\,
\int_0^1 dx \int \frac{d^2k_\perp}{2(2\pi)^3} \, \lf(1-x\rg) \\
\frac{\vect{k}_\perp^2 - \lf(1-x\rg)^2 M^2}
{\lf[\vect{k}_\perp^2 + (1-x)^2\,M^2 + x\,m_\pi^2\rg]^2}\ F_{\pi N}^2(x,k_\perp^2).
\label{spin}
\end{multline}
Numerical evaluation using our LFWFs gives
\begin{align}
\Delta u =  0.921, \qquad \Delta d = -0.424,
\end{align}
so that the fraction of the spin carried by the quarks in a bare nucleon is
\begin{align}
\Delta \Sigma = \Delta u + \Delta d = 0.497.
\end{align}
Using Eq.~\eqref{all} and the results
\begin{align}
Z_{N\pi} = 0.706, \qquad \Delta q_N^\pi = 0.0281,
\end{align}
implies that the nucleon spin sum, including the effects of the pion cloud is given by
\begin{align}
\Delta \Sigma_\pi = 0.365.  
\end{align}
In contrast with previous work, the term $\Delta q^\pi_N$ is greater that zero. This results from our relativistic treatment, and 
numerically arises from the cancellation of the two terms in the numerator of the integrand appearing in \eq{spin}. 
This is a small effect.
The value 0.365 is in good agreement with the central value 0.366 obtained in the global analysis 
(using $x_{min}=0.001$ ) of Ref.~\cite{deFlorian:2008mr}.
Future measurements made at higher energies may reduce this central value. However, the present agreement 
is very good, considering that the model wave function has no gluons.

\section{summary and discussion}
\label{sec:sum}
The main point of our work is to show that 
it is possible to construct a constituent quark model -- capable of reproducing the measured electromagnetic form factors --
in which the quark spin content of the nucleon is in qualitative agreement with experiment. This phenomenology is achieved by using
relativistically moving quarks, immersed in a cloud of pions. There are several  possible improvements to the model:
including more pionic terms, increasing the flexibility of the guess for the wave functions given in Eq.~\eqref{phi}, 
improving the treatment of the pion-nucleon vertex along the lines suggested by~\cite{Alberg:2012wr}, including 
the effect of  intermediate $\Delta$-baryons 
in the pion cloud contribution and  so on. While the present model is not likely to be the final word on the subject,  it 
does show that the quark model, with suitable obvious  modifications from the original non-relativistic, pion cloud-free 
version  does survive the ``proton spin crisis" in 
a manner very similar to that  previously noted~\cite{Myhrer:2007cf,Thomas:2008ga}.  Future refinements and tests of the model
depend on the ability of experimentalists to  make improved measurements.

\section*{Acknowledgements}
This work has been partially supported by US DOE Grant No. DE-FG02-97ER-41014
and by the Australian Research Council through FL0992247 (ICC).


\end{document}